# The Road to the Ideal Stent: A Review of Stent Design Optimisation Methods, Findings, and Opportunities

*A Kapoor\* [a], N Jepson [b,c], N W Bressloff [d], P H Loh [e,f], T Ray [g], S Beier [a]*

[a] *School of Mechanical and Manufacturing Engineering, University of New South Wales, High St. Sydney, Australia*

[b] *Prince of Wales Clinical School of Medicine, 18 High St, University of New South Wales, Sydney, Australia*

[c] *Prince of Wales Hospital, 320-346, Barker Street, Randwick, Sydney Australia*

[d] *University of Leeds, LS2 9JT, United Kingdom*

[e] *Department of Cardiology, National University Health Centre, National University Health System, Singapore*

[f] *Yong Loo Lin School of Medicine, National University of Singapore, Singapore*

[g] *School of Engineering and Information Technology, University of New South Wales, Canberra, Australia*

\*Corresponding Author

**Abstract**

Coronary stent designs have undergone significant transformations in geometry, materials, and drug elution coatings, contributing to the continuous improvement of stenting success over recent decades. However, the increasing use of percutaneous coronary intervention techniques on complex coronary artery disease anatomy continues to be a challenge and pushes the boundary to improve stent designs. Design optimisation techniques especially are a unique set of tools used to assess and balance competing design objectives, thus unlocking the capacity to maximise the performance of stents. This review provides a brief history of metallic and bioresorbable stent design evolution, before exploring the latest developments in performance metrics and design optimisation techniques in detail. This includes insights on different contemporary stent designs, structural and haemodynamic performance metrics, shape and topology representation, and optimisation along with the use of surrogates to deal with the underlying computationally expensive nature of the problem. Finally, an exploration of current key gaps and future possibilities is provided that includes hybrid optimisation of clinically relevant metrics, non-geometric variables such as material properties, and the possibility of personalised stenting devices.

# 1    Introduction

Coronary artery disease stands as the foremost cause of both mortality and morbidity on a global scale [1, 2]. This condition arises when the coronary arteries become narrowed due to plaque accumulation, thus restricting the supply of oxygenated and nutrient-rich blood to the heart muscle. As an established and effective treatment, Percutaneous Coronary Intervention (PCI) using stents offers a solution for reopening these narrowed arteries with an estimated two million annual coronary stent implants world-wide [3]. In fact, for many coronary artery disease patterns and even a large number of patient groups, PCI with stenting is a preferred interventional method as it is less invasive and has lower associated risk compared to the interventional alternative, bypass surgery.

The success of stenting is determined by considering both procedural and clinical factors. These factors include deliverability, which refers to the ease and accuracy of stent placement. Additionally, the effectiveness of stenting is measured by the minimisation or absence of In-Stent Restenosis (ISR), a condition characterized by excessive re-narrowing of the treated artery. Moreover, another critical aspect is the prevention of Stent Thrombosis (ST), a condition where local blood clot formation leads to sudden arterial occlusion, potentially causing myocardial infarction (heart attack) and even sudden death. Stent design improvements have been a key driver for increasing stenting success over the years. Stents were developed for their first clinical use as Bare Metal Stents (BMS) with good scaffolding support, yet poor deliverability and high ISR rates [4]. Later, two subsequent generations of Drug-Eluting Stents (DES) largely replaced BMS due to better deliverability and improved clinical outcomes, achieving revascularisation rates as low as 3% and late ST rate of 0.6% today (Figure 1). A parallel pursuit presented Bioresorbable Scaffolds (BRS) of dissolvable polymers or metals, in an attempt to advance this technology further by dissolving into the blood stream after successful vessel reopening [5]. Yet, BRS were never routinely used in clinical practice due to a potentially increased risk of late clinical failure compared to DES [6], but clinical research on improved BRS designs is still ongoing.

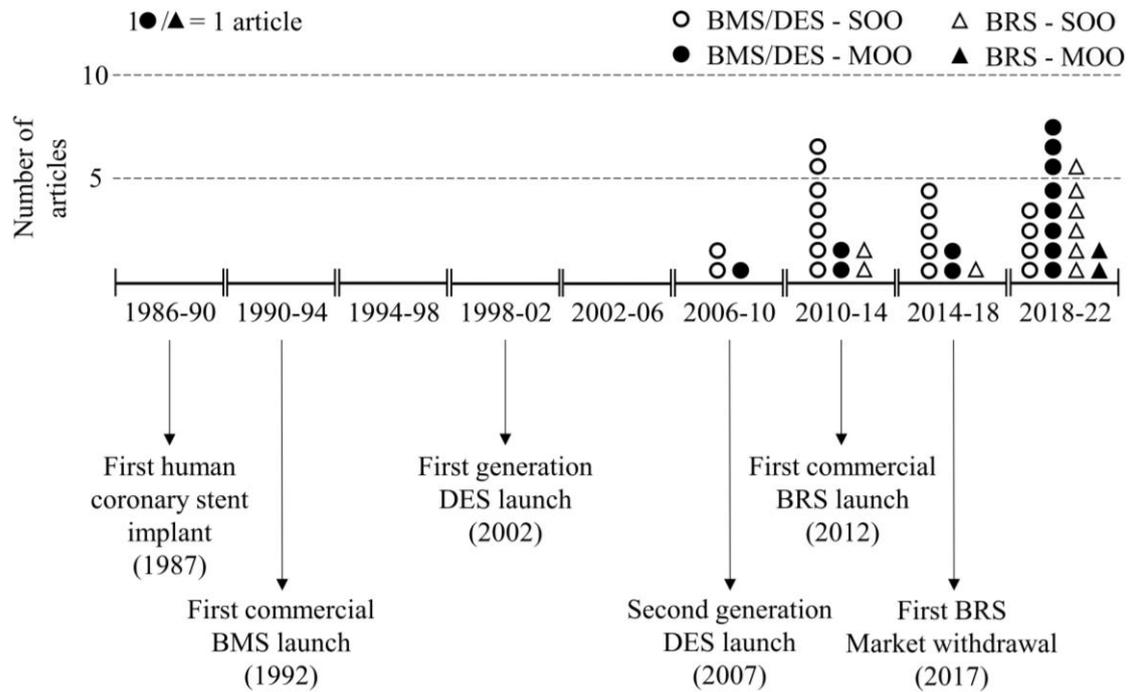

Figure 1: Number of articles published (top) on BMS / DES (circle), and BRS (triangle) concerning either single- (hollow) or multiple-objective optimisation (filled) studies in successive four-year intervals from 1986 till 2022, and milestones of coronary stent development (bottom). *Bare Metal Stent (BMS), Drug Eluting Stent (DES), Bioresorbable Scaffold (BRS), Single Objective Optimisation (SOO), Multi-Objective Optimisation (MOO).*

Optimisation has supported and enhanced these improvements of stent designs in the past fifteen years (Figure 1). Typically, optimisation algorithms can either optimise just one metric via Single Objective Optimisation (SOO) or several competing performance metrics in Multi-Objective Optimisation (MOO). Both techniques commonly use population based stochastic optimisation algorithms – techniques that use randomness and a set of initial points in a pre-defined design space to search for optimised solutions [7]. Since the earliest published work of systematic, computational optimisation of stent design, most studies have employed SOO but with an increasing tendency to use MOO in recent years.

The use of these optimisation techniques in relation to improving coronary stent design forms the focus of this review. Specifically, we first discuss current coronary stent designs and their design attributes, then their performance criteria, across structural, haemodynamic, and drug elution aspects including measurable indicators in the context of clinical outcomes. This is followed by a detailed description of size, shape and topology representation and optimisation schemes, describing stent design features as variables, before detailing the respective optimisation algorithms and their associated surrogate models. We then explore various testing methodologies of the optimised designs, and finally, we discuss research opportunities and promising future trends in the field, especially highlighting the potential of hybrid optimisation and examine the importance of clinically relevant metrics and non-geometric properties for future studies. Finally, we discuss the possibility of personalised stenting in future, providing an overall outlook from both engineering and clinical perspectives.

> *In silico* testing refers to the computational simulations performed to evaluate stent performance.
> *In vitro* tests involve experimental evaluations of stent performance on the 'bench-top', outside of a living organism.
> *In vivo* tests are performed in living organisms to observe the clinical effects of deployed stents.
> **Independent ring design** refers to the set of stent designs where the ring structures are independent and connected to each other by connectors only.
> **Helix based stent design** refers to a set of stent designs where the ring structure forms a part of a continuous single or a double helix with connectors providing additional axial connections between the rings.
> **Stent design family** represents the set of all design variations from a baseline design that can be explored by a stent design optimisation algorithm.
> **Surrogates** are approximate models of objectives/constraints that are used in lieu of expensive objective and constraint evaluation during the optimisation process. The term expensive refers to the cost which can be computational cost or cost of physical experiments.
> **Non-dominated solution** is a set of optimised design solutions, whereby no solution is better than any other member in all optimisation objectives.

## 2    Key developments from past to current contemporary coronary stent designs

Coronary stents, serving as vascular scaffolding implants over recent years, boast a remarkable history that has been examined in extensive reviews elsewhere [8, 9] and are thus only briefly presented here. We will refer to stents in terms of BMS, DES and BRS, whilst scaffolds exclusively apply to BRS.

The first deployment of the BMS [10] was a pivotal moment in coronary artery disease intervention, but the first version of BMS resulted in up to 20% ISR rates and thus many patients required vessel revascularisation [11]. This can be largely attributed due to their simplistic designs and especially due to their large struts. DES came into clinical use nearly a decade later, whereby novel anti-proliferative drug coatings successfully minimised ISR from 17 to around 4% [12]. However, within 1-2 years after implantation, first-generation DES resulted in alarming incidences of very late ST [13]. This sparked significant controversy within the cardiology community with some arguing that BMS are comparatively safer than first generation DES due to lower late thrombosis rates. Within 5-years, however, second-generation DES entered the clinical realm around 2007, with a number of advances: cobalt, chromium and platinum alloys instead of stainless steel material enabled the use of thinner struts (80-90 $\mu m$ compared to 132-140 $\mu m$ previously [14]) which resulted in reduced blood flow disturbance around stent struts, better biocompatible coatings, and less toxic anti-proliferative drugs [15]. In recent years, DES with ultrathin struts of <70 $\mu m$ and bioresorbable polymer coatings may have contributed to reduce ISR further with less than 3% need for revascularisation (from 4% previously) for new coronary lesions excluding bifurcations and chronic total occlusion [16]. Late ST was successfully reduced from 2.5 to 0.6% [17], yet early ST risk for even latest generation of DES still necessitates the use of dual anti-platelet therapy (DAPT) or blood thinners. The recommended duration of the medication continues to evolve as a balance has to be achieved between the clinical outcome including minimisation of DES-related early ST and bleeding complications, reduced patient wellbeing and significant financial burden from the ongoing medication [18, 19].

Modern DES structures commonly used clinically are comprised by three distinct design features: struts, crowns and connectors, whereby multiple struts and crowns join to form a ring. Successive rings are connected using dedicated connectors. Moderns DES currently in clinical use commonly have an open cell-design comprising rings and connectors with different shapes (Table 1). Rectangular struts with a thickness between 74-90 $\mu m$ are the current industry standard with exceptions of two stents: the circular cross-section strutted Resolute Onyx (Medtronic Vascular) and the 60 $\mu m$ ultrathin strutted Orsiro (Biotronic Inc.). Similarly, most stent rings are connected to each other through connectors while the double helix Orsiro stent [20], and single helix Resoute Onyx stent [21] are the only stents where rings are pre-connected through the helix in addition to the S type connectors (Orsio) or spot welding of adjacent crowns (Resolute Onyx). The most common DES material is cobalt chromium (CoCr), and again only two commercial stents differ: the Synergy (Boston Scientific Corp.) stent made of platinum chromium (PtCr), and Resolute Onyx with an inner platinum iridium (PtIr) core covered with a CoCr shell. As for the stent coating, earlier DES had either paclitaxel [22], or sirolimus-eluting coatings [12], whereby the latter has been established as superior in regards to both ISR and ST outcomes [23], and therefore either sirolimus or sirolimus-analogue drug coatings are commonly used. Further details on current stent design issues and future possibilities of DES can be found elsewhere [24]. In addition to stent design, different base materials and stent manufacturing techniques affect the clinical performance of the stents including the ISR and ST. However, these topics are outside the scope of the current work and readers are suggested to consult the other review articles for further reading [25-27].

Table 1: Design details of commercially available Drug-Eluting Stents (DES) approved by US Food and Drug Administration [20, 24, 28-37].

| Stent | Xience Sierra/ Skypoint | Resolute Onyx / Onyx Frontier | Synergy/ Synergy XD | Orsiro | Elunir | Slender |
|---|---|---|---|---|---|---|
| **Manufacturer** | Abbot Vascular | Medtronic Vascular | Boston Scientific Corp. | Biotronic Inc. | Medinol Ltd. | Svelte Medical Systems, Inc. |
| **Design type** | Independent ring[1] | Single helix[2] (single wire) | Independent ring[1] | Double helix[2] | Independent ring[1] | Independent ring[1] |
| **Connector type** | U type (peak to valley) | No connector – locally laser fused rings (peak to peak) | Straight (peak to peak) | S type (strut mid to mid) | S type (peak to peak) | S type |
| **Material** | CoCr | CoCr shell with PtIr core | PtCr | CoCr | CoCr | CoCr |
| **Drug coating** | Everolimus | Zotralimus | Everolimus | Sirolimus | Ridaforolimus | Sirolimus |
| **Drug dose** | $100\ \mu g/cm^2$ | $160\ \mu g/cm^2$ | $100\ \mu g/cm^2$ | $144\ \mu g/cm^2$ | $110\ \mu g/cm^2$ | $213\ \mu g/cm^2$ |
| **Coating type** | Conformal - Permanent Polymer | Conformal - Permanent Polymer | Abluminal - Bioabsorbable | Conformal Bioabsorbable | Conformal Permanent | Conformal Bioabsorbable |
| **Cross-section** | Rectangular | Circular | Rectangular | Rectangular | Rectangular | Rectangular |
| **Available stent diameters (mm)** | 2.25, 2.5, 2.75, 3.0, 3.25, 3.5, 4.0 | 2.0, 2.25, 2.5, 2.75, 3.0 3.5, 4.0, 4.5, 5.0 | 2.25, 2.5, 2.75, 3.0, 3.5, 4.0, 4.5, 5.0 | 2.25, 2.5, 2.75, 3.0, 3.5, 4.0 | 2.5, 2.75, 3.0, 3.5, 4 | 2.25, 2.5, 2.75, 3.0, 3.5, 4.0, |
| **Available stent lengths (mm)** | 8, 12, 15, 18, 23, 28, 33, 38 | 8, 12, 15, 18, 22, 26, 30, 34, 38 | 8, 12, 16, 20, 24, 28, 32, 38 | 9, 13, 15, 18, 22, 26, 30, 35, 40 | 8, 12, 15, 17, 20, 24, 28, 33 | 8, 13, 18, 23, 28, 33, 38 |
| **Strut thickness ($\mu m$)** | 81 | 81 | 74 – 81 (Diameter 2.25-2.75mm: 74; Diameter 3.0-3.5mm: 79; Diameter 4.0mm: 81) | 60 - 80 (Diameter 2.25-3mm: 60; Diameter 3.5, 4mm: 80) | 87 | 81 – 84 (Diameter 2.25-3mm: 81; Diameter 3.5, 4mm: 84) |

*Cobalt Chromium (CoCr), Platinum Iridium (PtIr), Platinum Chromium (PtCr)*

---

[1] Independent ring design refers to the stents where the ring structures are independent and connected to each other by connectors only.

[2] Helix design refers to stents where the ring structure forms a part of a continuous single or a double helix with connectors providing additional axial connections between the rings.

An ongoing challenge in improving clinical performance derives from the fact that DES are permanent implants, concerns remain about long-term issues such as abnormal vasomotion, risk of late stent fracture, late malapposition (between six to nine months [38]), and ongoing risk of stent failure [39]. To address these issues, BRS were introduced to clinical use in 2012 on the premise that vascular scaffoldings are only required during the early phases of coronary arterial remodelling following the scaffold implantation by preventing vessel recoil and early re-narrowing. Unlike BMS or DES, these scaffolds comprise of biodegradable polymers or metals which, as the name suggests, completely breakdown within two to three years. However, initial BRS were much thicker than DES (~150-170μm) to compensate for the weaker material strength [40], whilst retaining scaffolding capacity. While BRS were considered an extremely promising development, longer-term trials in 2017 showed sobering results with significantly higher ST rates of up to 2.5% within three years (compared to 0.6% for the latest DES) [41] leading to the discontinuation of the product by the manufacturer Abbott Vascular (2017). Since then, the safety perception of BRS has diminished, however, research continues to overcome the remaining drawbacks of stents as the most commonly used medical device to date [5, 42, 43].

Although the first generation Absorb BRS (Abbott Vascular, USA) had disappointing long-term outcomes, more than twenty new scaffolds are currently being developed to overcome the initial shortcomings [43]. For Poly-L-Lactic Acid (PLLA) polymer-based scaffolds, strut thickness in new designs has been reduced from approximately 150 $\mu m$ to 95 $\mu m$ in Arteriosorb (Arterius Ltd., UK) and 100 $\mu m$ in MeRes100 (Merryl Life Sciences, India). Innovative design with smaller cells at the stent centre and new manufacturing processes such as melt processing and die drawing were used to improve the radial strength while reducing strut thickness in Arterisorb [44]. In addition to PLLA, a thin strut BRS - Fantom Encore (REVA Medical, USA) based on a proprietary polymer Thyrocare with strut thickness of 95-115 $\mu m$ was recently developed (2018). Although most scaffolds in development are polymer based, a few designs use metals due to their inherently high material strength. Magnesium based BRS Magamris (Biotronik AG, Switzerland) with a strut thickness 150 $\mu m$ [45], and iron based scaffold IBS (Lifetech Scientific, China) with strut thickness 70 $\mu m$ are currently in human trials. These continuous developments in thin strut BRS technologies have provided the hope for a BRS led future of stenting technology.

It should be noted that stents - as an arterial support device - are used for reopening of occluded vessels at multiple locations other than the coronary arteries of course. These include the peripheral arteries like femoral and iliac arteries [46], ureters [47], or tracheal and bronchial airways [48]. A majority of these occlusions are treated with self-expandable stents made with super-elastic Nickel-Titanium alloy "Nitinol" in contrast to the balloon expandable coronary stents, and thus have different designs, deployment and deformation mechanism, and structural mechanics [46]. Here, we focus on development of coronary stent designs and their performance metrics, however, the optimisation methodologies presented can be generalised across any stents.

## 3  Performance criteria of coronary stents

Coronary stenting success is influenced by both acute procedural and long-term clinical outcomes. Procedural outcomes include stent deliverability, immediate increase in luminal cross-sectional area, the lack of arterial injury or immediate thrombus formation, and expansion quality of the stent. Long-term clinical success is indicated by the absence of ISR, late ST, and relevant clinical events including the requirement for repeat PCI.

Multiple engineering metrics have been defined that act as markers for stenting success across both immediate/ procedural and long-term considerations. It is important to note that stent performance must fulfil two sets of functions. First, structural support of the artery must be achieved with minimal vessel injury and minimal malapposition. Secondly, the minimal obstruction or alternation of the surrounding blood flow environment. Thus, the key performance indicators for balloon-expandable coronary stents can be classified overall into 1) structural, 2) haemodynamic, and 3) combined metrics (Figure *2*, Table *2*). Structural metrics, prominently considered in stent optimisation, explore parameters related to the structural response of a stent and arteries during and after stent deployment [4]. Haemodynamic factors consider the fluid dynamic forces at the arterial wall as a result of the changes in blood flow due to the stent, known to promote ISR in some instances [49]. Combined metrics include factors such as tissue prolapse and drug elution, which are in part related to both structural and haemodynamic effects. The expected value of the performance metrics for the open cell contemporary stent designs is provided in Table *2*.

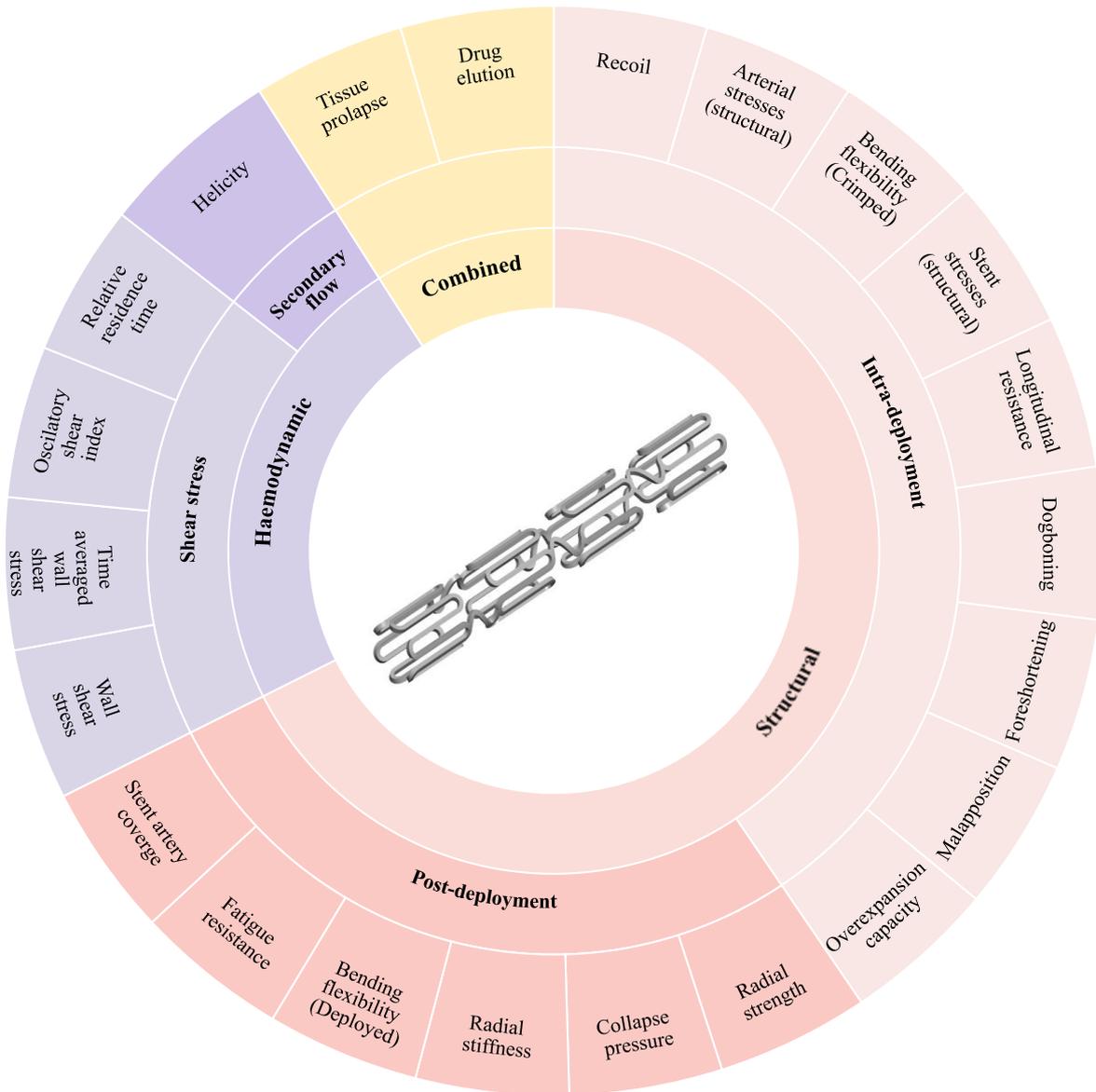

Figure 2: Key performance indicators for stent design including structural – deployment (light red) and post-deployment metrics (red), haemodynamic – shear stress (light purple) and secondary flow (purple) metrics, and combined performance metrics (yellow). A generic stent design depicted at the centre.

Table 2: Definitions, mathematical formulae, and expected values of key performance indicators for contemporary open-cell coronary stent design, including structural deployment (light red) and post-deployment metrics (red), haemodynamic shear stress (light purple) and secondary flow (purple) metrics, and combined performance metrics (yellow).

| Metric type | Performance Measure | Definition | Numerical Formula | Expected Values[1,2] | Description |
|---|---|---|---|---|---|
| **Structural Intra-deployment** | Stent recoil (%) | Reduction of stent diameter after balloon removal during deployment | $\left(1 - \dfrac{Diameter_{final}}{Diameter_{inflated}}\right) \times 100$ | 1.45 – 4.43 % [50] | $Diameter_{inflated}$ and $Diameter_{final}$ are the outer diameter of stent when the delivery balloon is fully inflated and after deflation respectively [51] |
| | Arterial stresses (structural) | Volume average stress developed in the artery during stent deployment | $\dfrac{\int_V \sigma dV}{\int_V dV}$ | 0.021-0.049 MPa [50] | $\sigma$ is the circumferential or von mises stress integrated over affected artery volume, $V$ [53, 54] |
| | Bending flexibility (crimped) | Ability of crimped stent to undergo bending deformation in a tortuous vessel path | $\int_0^{\kappa_l} M d\kappa$ | 0.012 – 0.064 N.Rad [50] | $M$ and $\kappa$ are bending moment and curvature in a bending test respectively. $\kappa_l$ is the user determined curvature limit [53, 54] Lower value indicates higher flexibility. |
| | Stent stresses (structural) | Structural stress imparted on the stent during deployment | $\dfrac{\int_V \sigma dV}{\int_V dV}$ | -- | $\sigma$ is the circumferential or von mises stress integrated over stent volume, $V$ [55] |
| | Longitudinal resistance | Ability of deployed stent to resist deformation under an axial compressive load | $\int_0^{\gamma_d} F(\gamma) d\gamma$ | 0.38 – 0.88 N [57] | F is the compressive force on stent and $\gamma$ is ratio of initial and final length of stent [54] |
| | Dogboning | Slower expansion of stent middle as compared to end locations during deployment | $\dfrac{r_{distal} - r_{central}}{r_{distal}}$ | 0.44 – 0.62 [54, 58] | $r_{distal}$ and $r_{central}$ are the radius at the ends and middle of stent respectively [4] |
| | Foreshortening (%) | Decrease in axial length of stent during expansion | $\left(\dfrac{L_{final} - L_{initial}}{L_{initial}}\right) \times 100$ | -- | $L_{initial}$ and $L_{final}$ are the lengths of the stent in crimped and deployed condition respectively. [4, 59] |
| **Structural** Post-deployment | Malapposition | Incomplete apposition of stent struts to arterial walls | $\frac{\int_A SM dA}{\int_A dA}$ | 0 mm | $SM$ is the Euclidian distance between a point on outer surface of stent with area $dA$ and its projection to the lumen [61] |
| | Overexpansion capacity | The capacity of the proximal segment of stent to be expanded above its labelled nominal diameter [63] | -- | 4 - 4.4 mm [64] | -- |
| | Radial strength | Maximum compressive load bearing capacity of a deployed stent | -- | 0.101 - 0.632 N/mm [50] | The evaluation requires a Segmented head or sling type setup [65] |
| | Collapse pressure | The radial compressive load that causes buckling in a deployed stent | -- | -- | The evaluation requires a hydraulic/pneumatic type setup [65] |
| | Radial stiffness | Rate of change of compressive force with radial deflection of a deployed stent [66] | -- | -- | -- |
| | Bending flexibility (deployed) | Ability of deployed stent structure to deform under bending force or moment | $\frac{\delta}{P}$ or $\frac{k}{M}$ | -- | $P$ and $\delta$ are the force and deflection in linear range of fixed span three point bending test. $M$ and $\kappa$ are midspan bending moment and curvature in linear range of variable span three point bending test [67] |
| | Fatigue resistance | Resistance of deployed stent against fatigue fracture due to pulsating load from heart rhythm | $\frac{1}{FSF} = \frac{\sigma_m}{\sigma_{uts}} + \frac{\sigma_a}{\sigma_N}$ | -- | $FSF$ is the fatigue safety factor. $\sigma_m$ and $\sigma_a$ are mean and alternating stresses respectively. $\sigma_{uts}$ and $\sigma_N$ are ultimate tensile strength and endurance limit of material respectively. $\frac{1}{FSF} < 1$ indicates safe operating zone [54] |
| | Stent artery coverage | Area of the artery covered by stent surface | $\frac{S_{stent}}{S_0} \times 100$ | ~ 13 % [68] | $S_{stent}$ is the external surface area of the crimped stent $S_0$ is the internal surface area of the artery covered by stent [60] |

| | | Metric | Description | Formula | Expected range | Notation |
|---|---|---|---|---|---|---|
| **Haemodynamic** Shear Stress | | Wall Shear Stress (WSS) | Shear stress applied on arterial wall by blood flow | $n.\overrightarrow{\tau_{ij}}$ | 0.4 Pa [69] -3 Pa [70] | $n$ is the normal vector to artery surface $\overrightarrow{\tau_{ij}}$ is the fluid viscous stress tensor [71] |
| | | Time Average WSS (TAWSS) | Average wall shear stress in a cycle of transient haemodynamic simulation | $\frac{1}{T}\int_0^T |WSS|dt$ | 0.4 Pa [69] - 3 Pa [70] | T is the time duration of a cardiac cycle [71] |
| | | Oscillatory Shear Index (OSI) | A measure of the oscillatory nature of wall shear stress | $0.5\left(1-\frac{\left|\int_0^T WSSdt\right|}{\int_0^T |WSS|dt}\right)$ | < 0.2 [73] | T is the time duration of a cardiac cycle [71] |
| | | Relative residence time (RRT) | A combined metric of OSI and TAWSS indicating the residence time of blood particles near wall | $\frac{1}{TAWSS(1-2\times OSI)}$ | < 4.17 $Pa^{-1}$ [73] | -- |
| **Haemodynamic** Secondary flow | | Helicity | A measure indicating the extent to which the velocity field coils and wrap around each other | $LNH = \frac{V(s,t).\omega(s,t)}{|V(s,t)||\omega(s,t)|}$ | -- | LNH is local normalised helicity V is the velocity vector $\omega$ is the vorticity vector [74] |
| **Combined** | | Tissue prolapse | The encroachment of the free luminal area between stent struts by an inward bulging tissue segment | $PI = area\ of\ largest\ inscribable\ convex\ quadilateral\ in\ deployed\ stent$ | $0.639 - 1.311\ mm^2$ [50] | PI is the prolapse index [76] |
| | | Stent drug elution | Amount and distribution of antiproliferative drug in artery | -- | -- | -- |

---

[1] The expected values of the structural metrics are provided for contemporary open cell metallic coronary stent designs with 3mm nominal diameter.
[2] The expected range of haemodynamic metrics are linked with physiological blood flow and is independent of stent design.

### 3.1 Structural

Structural performance metrics are the measures that describe the mechanical behaviour of the stent and vessel and can be broadly classified into intra- and post-deployment. Both play an important role in determining the overall stenting success and are thus critical in stent design optimisation research (Table 2).

Intra-deployment of a stent requires consideration of (i) vascular injury, (ii) deliverability, (iii) structural and longitudinal integrity, and (iv) expansion quality.

(i) Two main measures of vascular injury are radial recoil of the stent during deployment and imposed mechanical stresses on the arterial wall. A higher recoil tendency of a stent may require overexpansion compared to the vessel calibre, yet this increases the risk of vessel injury marked by the destruction, in part or completely, of the endothelial cells in the implant area, which promotes ISR [77]. Arterial stresses are an approximator for vascular damage [78]. It is important to note that both recoil and arterial stresses are important measures with a conflicting relationship, i.e., thicker struts reduce radial recoil but increase arterial stresses [53, 54]. Therefore, optimisation should account for both [50, 53-55, 79, 80], yet some work includes radial recoil [81-84], or arterial stresses [85, 86] separately.

(ii) Stent deliverability is measured as the stents bending flexibility in a crimped state. High bending flexibility improves the deliverability of the stent relevant to navigate a tortuous vessel path to reach the target lesion site. While the ASTM standards [67] recommend testing of bending flexibility of the complete deployment system (including crimped stent on balloon and guidewire), design optimisation studies performed simplified assessments only considering the stent itself [50, 79].

(iii) A stent's structural integrity concerns the stresses within the stent structure due to deployment, studied in only a limited number of optimisation studies to date [55, 87-89]. Longitudinal integrity is relevant for expanded stent's resistance to compressive axial forces, especially during post-dilation procedures of PCI, and has been studied both numerically [90] and experimentally [91, 92], and recently included in optimisation research [50, 93].

(iv) The stents expansion quality is determined by foreshortening, dogboning, malapposition, and overexpansion capacity. Foreshortening and dogboning depend on the stent design only and have been regularly used in stent design optimisation studies [54, 60, 82, 89, 94]. Malapposition is a composite metric that depends on multiple factors including the stent design, disease geometry, disease mechanical property, and stenting technique. Malapposed struts can act as thrombosis sites and therefore acute malapposition (> 0.4 mm) observed during stent implantation is corrected with post dilation procedures [95], which may in-turn cause overexpansion and vascular injury, highlighting the required balance between different metrics. Malapposition has been considered for optimisation of stent dilation procedure but not for stent design [96]. Overexpansion capacity is an important metric for selection of stents in complex PCI procedures such as bifurcation stenting, which has been evaluated *in vitro* for most commercial stent designs [63, 64]. While overexpansion capacity is an important clinical metrics, it has not been used as objectives in stent design optimisation to date.

Post-deployment performance measures include the (i) continuous scaffolding support to artery, (ii) bending flexibility (affects artery shape) [67], (iii) stent fatigue due to cyclic loading from heart rhythm, and (iv) arterial coverage of stent structure.

(i) Ongoing scaffolding support is a key structural consideration, especially for BRS which are composed of comparatively weaker and degradable base material and therefore extensively used in BRS design optimisation [60, 83, 97, 98]. Scaffolding support can be evaluated by radial strength [50, 83], or collapse pressure [60] defined as per ASTM standards [65], unlike the comparatively outdated radial stent stiffness metric [66, 97].

(ii) As for bending flexibility, deployed stents with low bending flexibility can permanently straighten the artery shape, causing long-term injury especially at the stent ends [98-100].

(iii) Stent fatigue due to cyclic loading from the heart rhythm may cause fractures, which can ultimately lead to thrombosis. Unlike structural stresses that cause instantaneous fracture during stent expansion, fatigue is a long-term phenomenon. For BRS, this may be a major concern due to their overall lower structural strength [40], yet this has not been studied in optimisation research before. For BMS or DES comprised of alloys with higher structural integrity (cobalt/platinum-based alloys), this is a minor concern due to low incident rates [54, 101, 102].

(iv) Finally, arterial coverage, also known as stent footprint, describes the stent surface area coverage of the artery and is associated to ST incident rates [103, 104]. BRS have significantly higher footprint (~30% vs 13% for DES [68]) due to lower material strengths and increased geometric dimensions, and thus this measure has been used exclusively for BRS optimisation [60, 83].

### 3.2 Haemodynamic

Haemodynamic stenting metrics and their clinical relevance have been previously reviewed in depth [49, 105], with *in silico* investigations using 3D Computational Fluid Dynamics (CFD) being most common [72, 75, 106, 107]. However some *in vivo* work in animals [108], and humans [109-111] exist. Instantaneous Wall Shear Stress (WSS) is a well-established indicator of endothelial cell effects directly linked to ISR [112], and thus WSS [86, 113] or its cardiac cycle Time-Average (TAWSS) [66, 79, 114, 115] have often been used as haemodynamic objectives. One study also included the Oscillatory Shear Index (OSI), and Relative Residence Time (RRT) to capture oscillatory blood flow dynamics for ascertaining the link between restenosis and varying stent positioning during implant [108]. Haemodynamic effects are continuous, yet optimisation methods use adversity thresholds, which are not fully agreed upon in the literature (e.g. 0.5 Pa [79, 86] vs 1 Pa [49, 66] for both WSS and TAWSS). 2D secondary flow characteristics, including flow vorticity, recirculation zone length, reattachment length, and swirl have been considered in the past for optimisation [113, 116]. The increasing recognition of their 3D counterparts such as helicity [117] yields the notion that considering 3D secondary flow metrics as an optimisation objective may be an opportunity in the field.

### 3.3 Combined performance metrics

Some metrics combine structural and haemodynamic aspects of stenting, including tissue prolapse and drug elution. For tissue prolapse, whilst the degree of prolapse is a structural outcome related to the stent design [76, 118] and target lesion characteristics (e.g. higher prolapse in soft lipid atheroma versus harder calcified plaques [119]), it has profound haemodynamic implications due to its fluid dynamic effect on the boundary layer region [120, 121]. Whilst studies previously considered prolapse both directly [50, 54], and indirectly (within the haemodynamic performance objective) [79, 86], no study has evaluated prolapse for the latest generation of stent designs to date. Similarly, drug elution, a critical factor in DES performance [122], is based on structural design and deployment aspects as well as fluid dynamic distribution considerations. The drug elution characteristics [123] and distribution in an artery [53, 79] have been optimised before.

## 4  Stent optimisation

Stent design optimisation efforts generally aim to improve performance through an iterative framework comprising the problem definition, evaluation of objectives and constraints, applying the optimisation scheme that iteratively identifies possible design improvements/solutions, and finally the selection and testing of final designs (Figure 3). The problem definition phase involves identification of design variables based on different representation methodologies (size, shape or topology optimisation – Section 4.1), optimisation objectives (performance metrics – Section 0), constraints on the objectives or design variables, and initial population of designs using domain sampling methodologies. Thereafter, multiple iterations of design evaluation and optimisation phase occur in conjunction, propose, and evaluate new designs. The testing is performed on the final design selected after optimisation phase.

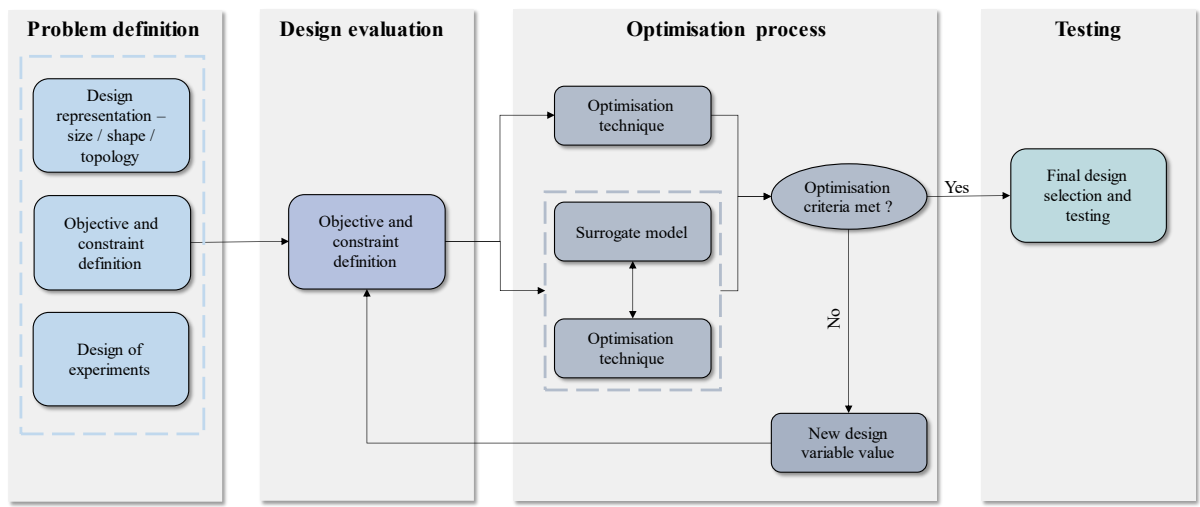

Figure 3: Stent design optimisation framework representing the problem definition (section 4.1), design evaluation (section 4.2), optimisation process (section 4.3), and final design testing (section 4.4) stages of the optimisation process. The problem definition phase involves identification of design variables using different representation methodologies (size, shape, or topology optimisation), optimisation objectives or constraints, and initial population of designs using domain sampling methodologies. The design evaluation phase involves estimation of objectives and constraints using numerical simulations techniques. This is followed by optimisation process where optimisation techniques are used (either independently or coupled with surrogate models) to suggest new designs. The suggested designs are then evaluated for objectives/constraints and the process is repeated unless the optimisation criteria are met and the final design is selected to undergo further testing.

### 4.1  Problem definition

In stent design optimisation, variables representing the stent design, the overall optimisation objectives, and their constraints need to be defined. Generally, objectives and constraints are domain specific, whereby objectives of stent optimisation are derived from clinically key measures of stents performance (Figure 2), and are either being minimised (e.g., least flow obstruction) or maximised (e.g., maximum radial strength). Constraints allow designers to set realistic limits of such objectives. For example, constraints can involve numerical bounds on optimisation objectives [60, 80, 89, 123] and other non-objective performance metrics [81], or a constraint for improvement of objective values above a baseline design [53]. The evaluation of candidate designs is resources extensive, and as a result optimisation algorithms are commonly embedded with surrogates or approximations to rapidly identify suitable solution. The initial surrogates are based on designs sampled across the variable space.

Common form of sampling includes ones based on Latin Hypercube Sampling (LHS) [124], central composite design, lp$\tau$, modified rectangular grid [94, 125], and Halton sampling [86, 126]. Among these, LHS strategy is most common [50, 60, 66, 80, 83, 84]. The stent optimisation design variables concerning geometry can be based on three different representation schemes, size, shape or topology optimisation.

### 4.1.1 Size and shape optimisation

Size and Shape optimisation are used in most stent optimisation studies. Size optimisation involves modifying the dimensions of various stent design features like strut, connector, or crown, whereas the shape optimisation involves changes in overall design of these features. Size optimisation can be performed independently [50, 80] or concurrently with shape variations in a single optimisation routine [66, 89]. In any size or shape optimisation exercise, parameterisation is critical as it defines the family of designs that can be explored by the underlying optimisation algorithm. Geometry-based parameters are the most common compared to parametrisation of manufacturing processes or material properties. For the former, a base geometry is defined using a small number of variables, typically between two and eight. The baseline geometry is often inspired by existing designs, and intelligent perturbations to the parametrised variables during optimisation allows for the exploration of a family of stent designs (such as helical or independent ring-connector base designs).

The geometric parametrisation of stent can be performed in three different base configurations - uncrimped, crimped, or deployed. The stents are manufactured in the uncrimped state, compressed on the delivery balloon and guidewire to the crimped state, and finally expanded in the artery to the deployed state. The most common base configuration for geometric parametrisation is the crimped state [53, 54, 60, 80, 98], followed by deployed [66, 86, 115] and uncrimped states [83, 88, 127]. Deployed configuration is a preferred base state for haemodynamic-focussed optimisation studies [66, 86], but it presents an idealised version of the stent's deployed geometry, which is different from the actual deployed configuration obtained by performing expansion of a crimped stent. Moreover, since the stent expansion from crimped to deployed is not simulated, it is likely that structural metrics are not accurately captured due to the lack of deployment related stress history. The initial crimping of DES is generally ignored due to their excellent material properties, unlike BRS in which the weaker base material is vulnerable to crimping damage [88]. Therefore, the uncrimped BRS are commonly used for parametrisation [83, 88].

Stent design is represented by multiple geometric design variables in optimisation (Figure 4, Table 3). The geometric parametrisation schemes includes different size variables related to the stent strut width [53], thickness [66], and length [60], connector width [55] and height [53], or crown radii [89]. Shape optimisation of stent is performed by altering the types of cross-section [66], centrelines [66, 101], number of connectors and rings [89], or alignment of rings within the stent [66] (Figure 4). The size optimisation variables are continuous, whereas the shape optimisation variables are generally discrete or categorical (Table 3). The bounds of design space for the geometric size and shape optimisation are typically derived from commercial stents [60, 66, 85], physical feasibility [101], or geometric limitations such as prevention of self-contact between struts [53, 54, 60]. A common method of defining a design space is to use different commercially available stents to decide the range of variable such as strut width and thickness (0.06 – 0.12 mm [66]), strut cross section (circular or rectangular [66]), or ring alignment (in- or out of- phase [66]). In other cases, bounds for the design space originate from physical feasibility to manufacture stents with varying strut thickness (0.03 – 0.15 mm) [101], prevent meshing constraints for low width structures ($<$ 0.065 mm) [54], or avoid self-contact between high width struts ($>$ 0.115 mm [54] or 0.17 mm [53]). Overall, the exact values of these constraints depends significantly on the baseline geometry of the stent chosen for parametrisation.

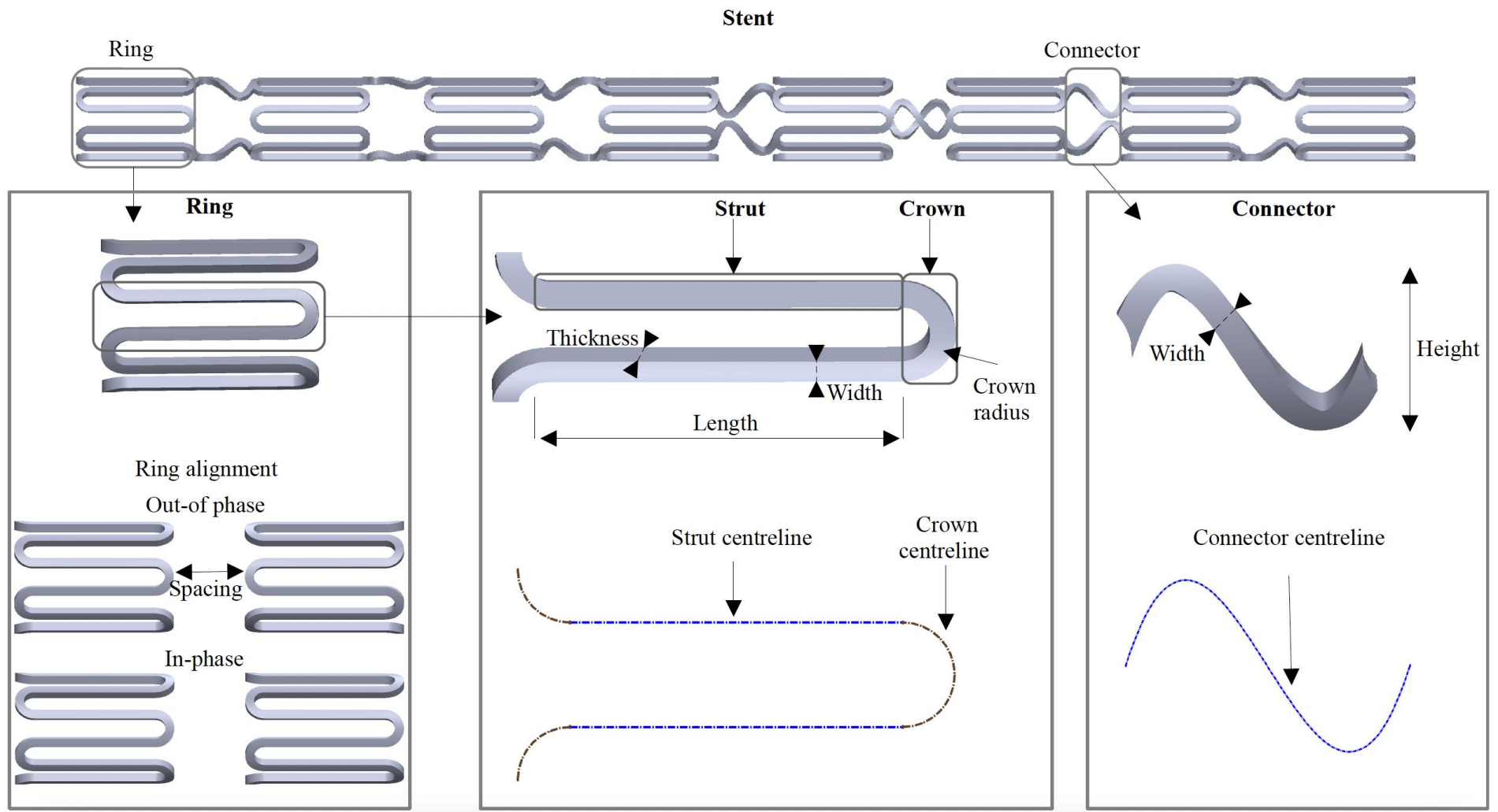

Figure 4: General parametrisation scheme for geometric design variables in stent design optimisation. A generic open-cell independent ring stent design with labelled design variables for the stent cell (left), strut (middle), and connector (right). The parametrisation scheme demonstrates the size optimisation variables (strut – width, thickness, length; connector – width, height; crown radius) and shape optimisation variables (strut – centreline, cross-section; connector - centreline, number per cell; ring – alignment, number per stent).

Table 3: Details of the geometric design variables including struts, connector, and cell of a generic coronary stent with the effect of each variable on the stent's performance metrics.

| Stent features | Design variable | Type | Range | Variable change | Performance Metrics | Beneficial effect | Adverse effect | Study reference |
|---|---|---|---|---|---|---|---|---|
| **Strut** | Width | Continuous | 0.05 – 0.18 mm | Increase | Arterial stresses, drug elution, longitudinal resistance, radial strength, recoil, arterial stresses | Drug elution, longitudinal resistance, radial strength, recoil | Arterial stresses | Pant et al. [53], Ribeiro et al. [54] |
| | Length | Continuous | 0.04-0.4 mm | Increase | Arterial stresses, foreshortening, recoil, tissue prolapse | Arterial stresses, foreshortening, tissue prolapse, | Recoil | Pant et al. [53], Ribeiro et al. [54], Blair et al. [60] |
| | Thickness | Continuous | 0.04 – 0.45 mm | Increase | Bending flexibility (deployed), collapse pressure, WSS, TAWSS | Collapse pressure | Bending flexibility (deployed), WSS, TAWSS | Gharleghi et al. [66], Blair et al. [60], Baradaran et al. [98], Beier et al [72] |
| | Cross section | Categorical | Rectangular, circular, triangular | Rectangular to: Triangular Circular | WSS TAWSS | WSS TAWSS | -- | Putra et al. [86] Gharleghi et al. [66] |
| | Centreline | Categorical | Slot, Sinusoidal, splines | -- | -- | -- | -- | Baradaran et al. [98], Clune et al. [101] |
| | Number per cell | Discrete | 16-24 | Increase | Foreshortening, stent stresses | Foreshortening, stent stresses | -- | Torki et al. [89] |
| **Crown** | Crown radius | Continuous | 0.1 – 0.5 mm | Increase | Foreshortening, stent stresses | Foreshortening | Stent stress | Torki et al. [89] |
| **Connector** | Width | Continuous | 0.05-0.20 mm | Increase | Bending flexibility (crimped), longitudinal resistance | Longitudinal resistance | Bending flexibility (crimped) | Shen et al. [93], Amirjani et al. [55], Tammareddi et al. [80] |
| | Height | Continuous | 0 – 1.9 mm | Increase | Bending flexibility (crimped), WSS | Bending flexibility (crimped) | WSS | Pant et al. [53, 79, 128] |
| | Centreline | Categorical | Straight, spline, S-shaped | Straight to spline/S-shaped | Stent fatigue | -- | Stent fatigue | Ormiston et al. [102], Gharleghi et al. [66] |
| | Number per cell | Categorical | 1, 2, 3 | Increase | Bending flexibility (crimped), longitudinal resistance | Longitudinal resistance | Bending flexibility (crimped) | Ormiston et al. [91], Gharleghi et al. [66] |
| **Ring** | Number per stent | Discrete | 10-18 | Increase | Foreshortening, stent stresses | -- | Foreshortening, stent stresses | Torki et al. [89] |
| | Alignment | Categorical | In-phase, out-of-phase | In-phase to out-of-phase | Longitudinal resistance | Longitudinal resistance | -- | Choudhary et al. [129], |
| | Spacing | Continuous | 0.83 – 3 mm | Increase | TAWSS, WSS | TAWSS, WSS | | Gharleghi et al. [66], Putra et al. [86], Beier et al [72] |

*Wall Shear Stress (WSS), Time-Averaged WSS (TAWSS)*

The effects of the design variables on structural, haemodynamic, and combined performance metrics are listed in Table 3. For many design variables, changes can lead to a mix of beneficial and adverse effects on the metrics of interest. As a result, the complex nature of the design problem requires optimisation across multiple objectives, several of which are likely to be in competition with others. The most comprehensive size optimisation study on structural aspects to date by Ribeiro and co-authors [50], considered three variables i.e. strut width, length, and equivalent link length, analysing arterial stresses, radial recoil, bending resistance (crimped), radial strength, prolapse index, and longitudinal resistance. The optimisation algorithm produced 21 different designs with improved performance relative to a baseline model based on the commercial Promus Element (Boston Scientific) design. Similarly, the most extensive parametrisation scheme including both size and shape parameters used seven different design variables by Gharleghi et al. [66], including a first-time utilisation of categorical variables such as cross-section and connector centreline, allowing vastly different stent designs within an independent ring type baseline design to be analysed.

The base geometry for parametrisation is either derived directly from a commercial design with the dimensions acquired through imaging [50, 54], provided by the manufacturer [80], obtained from the literature [79, 82]) or is a researcher generated theoretical design [86, 89, 98]. For commercial design-based optimisation studies, all stent geometries have an independent ring design. While the independent ring structure is clinically used in multiple contemporary stents such as Xience (Abbot Vascular) and Synergy, other helix based stent designs such as double-helix in Orsiro [20] and single-helix structure in Resolute Onyx [21] have not been optimised in known published work. Additionally, while optimisation studies have incorporated different features of commercial independent ring stents in their parametrisation scheme [66], no single study compares vastly different stent designs e.g. helix, double helix and independent ring in a single study, where each of these may require individual parametrisation. Although geometry-based design variables have predominately been used for stent design optimisation, limited studies exist concerning other aspects such as drug elution [123], the stents manufacturing process [81, 130], stent implantation [80], and material property [80]. For drug elution, a computational transport model based optimisation of initial drug concentration and drug release rate showed that paclitaxel drug performance is optimal if the drug release is very fast (few hours) or slow (several months), while sirolimus requires only slow drug release kinetics [123]. Stent manufacturing technique optimisation used micro-injection moulding parameters [81, 130] for reduced warpage (variation from desired part shape during injection moulding process), enabling improved mechanical properties. Finally, one study showed that including the uncertainty due to different balloon positioning and material yield strength (due to manufacturing variability) in optimisation led to designs with less optimal performance metrics than cases where the uncertainty effect was not included [80]. No work has considered material choice as a design variable in addition to the geometry-based design variables to date.

#### 4.1.2 Topology optimisation

Stent design can also be formulated as a topology optimisation problem, whereby the domain (either 2D if optimising a cell, or 3D) is discretised into small pixels/voxels. The distribution of the material over the domain is then sought by allocating material to the domain elements with the aim to improve the structural or haemodynamic metrics. In most topology optimisation studies, the stent structure is represented as axial and circumferential repetitions of 2D unit cells, whose topology is optimised [131-133]. Although rare, direct discretisation of 3D domain has also been performed [134]. As the algorithm aims to improve design objectives by directly filling material over the discretised 2D or 3D domain, topology optimisation results are far more generic (i.e., can search across multiple families of designs), and involve a much larger number of variables (~1,000 for 2D to 30,000 for 3D) [131-135]. Topology optimisation algorithms have been used to generate innovative designs such as the auxetic stents the increase in length at expansion [132, 133], and bistable stents that automatically transition from a contracted state to a stable expanded state by application of a specific radial force [131]. Comparatively, shape optimisation operates within a family of designs, thus resulting in similar designs which are less novel and considers fewer variables (commonly less than 8).

A major drawback in topology optimisation is related to the presence of a number of variables and inability to deal with multiple metrics. Till date, a maximum of three simplified objectives have been simultaneously optimised using this representation scheme [132]. In terms of performance metrics, for structural considerations, radial stiffness was considered as the single objective [131, 135, 136], while stiffness and auxetic property of the stent was concurrently optimised [133] using topology optimisation. This was later extended to multidisciplinary topology optimisation by including the haemodynamic objective fluid permeability to achieve stent designs with reduced blood flow disturbance while maintaining the desired auxetic property [132]. Therefore, only three simplified objectives have been simultaneously considered in topology optimisation till date. A detailed description of different topology optimisation approaches is provided elsewhere [137].

In summary, topology optimisation strategy is suited to generate initial "out-of-the-box" stent design concepts (e.g. auxetic [132, 133] and bistable [131] designs) using limited performance metrics in the initial design phase, while size/shape optimisation can be used in later half of design cycle to optimise large number of objectives once a baseline design has been established.

### 4.2 Design evaluation

The design evaluation for all representation methodologies (size, shape, and topology) involves the assessment of the structural, hemodynamic, or other objectives and constraints for different stent designs using computer simulation techniques such as Finite Element Method (FEM) and CFD. These simulations are also used to underpin the sensitivity analysis evaluations of performance metrics on different design variables. FEM is the standard numerical technique for evaluating structural objectives. 3D FEM is currently state-of-the-art [50, 60, 66, 79, 83], while 2D FEM was used to evaluate structural objectives in a few earlier studies [87, 88, 97]. In addition to the calculation of structural metrics, FEM has been used to evaluate drug transport [53, 123], manufacturing process [81, 130], and in one case, a haemodynamic objective [116]. In general, CFD is the preferred mode of haemodynamic objective evaluation. While 2D CFD [113] was used to assess haemodynamic objective in earlier studies, most recent studies use sophisticated 3D CFD assessments [66, 79, 86]. Similar to FEM, CFD was also utilised in one assessment to evaluate drug accumulation and distribution in an artery [79].

### 4.3 Optimisation process

The optimisation process iteratively proposes new designs for evaluation to improve selected performance metrics. The optimisation process requires the selection of two important attributes – optimisation technique/algorithms and surrogate models (Table 4).

Table 4: Coronary stent design optimisation techniques, algorithms, and surrogate models with their update strategies.

| Technique | Algorithm | Surrogate – update strategy | Study References |
|---|---|---|---|
| Single objective Optimisation | Sequential quadratic programming | Kriging – No update | Pant et al. [53] |
| | | Kriging – EI | Li et al. [81, 94, 125] |
| | Mesh adaptive direct search [138] | Kriging – Believer | Gundert et al. [114, 115] |
| | | Kriging – PI | Bozsak et al. [123] |
| | EGO [139] | Kriging – EI | Ribeiro et al. [54], Grogan et al. [97] |
| | Global response surface method [140] | Gaussian radial basis – EI | Chen et al. [88] |

| | Sequential least squares programming | Quadratic – Believer | Blair et al. [60] |
| --- | --- | --- | --- |
| | Simulated annealing [141] | Quadratic – Believer | Wu et al. [87] |
| | Differential evolution [142] | NS | Mazurkiewicz et al. [143] |
| | Solid isotropic material with penalisation | NS | Wu et al. [135], James and Waisman [131], Li et al. [134] |
| | Parametric level set method | NS | Xue et al. [132, 133] |
| **Multi objective optimisation** | Non-dominated sorting genetic algorithm – II [144] | Kriging – Believer | Gharleghi et al. [66], Pant et al. [79] |
| | | Kriging – EI hypervolume | Gharleghi et al. [66] |
| | | Kriging – No update | Clune et al. [101] |
| | | NS | Baradaran et al. [98] |
| | | Polynomial – No update | Li et al. [83] |
| | EGO [139] | Kriging – No update | Srinivas et al [113] |
| | | Kriging – EI Hypervolume | Putra et al [86], Ribeiro et al. [50] |
| | | Kriging – PI Hypervolume | Ribeiro et al. [50] |
| | | Kriging – ParEGO | Ribeiro et al. [50] |
| | | Kriging – S metric selection-based EGO Hypervolume | Ribeiro et al. [50] |
| | Particle swarm [145] | Cubic – No update | Tammareddi et al. [80] |
| | $\epsilon$ multi-objective [146] | NS | Blouza et al. [116] |

*Expected Improvement (EI), Probability of Improvement (PI), Efficient Global Optimisation (EGO), No Surrogate (NS)*

### 4.3.1 Optimisation techniques

Optimisation algorithms are typically used to identify new sampling locations in a given design space that are likely to lead to improved objective values, while satisfying the constraints of the problem. Surrogate models are likely to need updated designs at locations other than at predicted optimal locations, as a means for improving the accuracy of the model itself.

SOO minimises or maximises only one objective by modifying the design variables, is most commonly applied in stents. . The objective can either be a physical performance metric [54, 88, 97, 143], a linear combination of multiple metrics [60, 81, 84, 89, 123, 130], or a composite function of performance metrics [82]. Although the linear combination of multiple objectives to create a SOO problem is a popular method to deal with more than one performance metrics, it enforces predetermined constraints on the optimisation algorithm based on the values of weights in the linear combination. Single objective techniques have used different numerical optimisation algorithms such as simulated annealing [141], efficient global optimisation [139], mesh adaptive direct search [138], and differential evolutions [142] to perform optimisation (Table 4). A Deep Reinforcement Learning (DRL) based algorithm called Proximal Policy Optimisation was recently used in shape optimisation of flow diverter stents for intracranial aneurysms [147]. In spite of significant excitement in Machine Learning (ML) community, superiority of these deep learning based shape optimisation algorithms over classical genetic optimisation is still under debate [148], and they have not been used in coronary stent design optimisation yet.

MOO enables the simultaneous handling of multiple objectives to obtain a set of non-dominated optimised solutions [146], from which an appropriate design can then be selected [79, 80, 86, 101, 113]. MOO methods do not require prior weights and are better suited to deal with complex problems, which require trade-off solutions like stent design optimisation. The most commonly used multi-objective optimisation algorithms is the Non-dominated Sorting Genetic Algorithm – II (NSGA-II) [144], but particle swarm [145], epsilon multi-objective [146], and efficient global optimisation-based algorithms [139] have also been used in some studies (Table 4). Gradient based continuous optimisation techniques such as efficient global optimisation [139], and nature-inspired direct optimisation algorithms such as NSGA II [144], Particle swarm [145], and simulated annealing [141] have been used in optimisation of coronary stent design irrespective of the types of surrogate models or design representation methodologies (size, shape or topology optimisation). DRL based optimisers for general stenting applications are currently limited to SOO problems as performing MOO to generate evenly distributed non-dominated solution set is a topic of ongoing research [149].

### 4.3.2 Surrogate model

Optimisation techniques require hundreds or thousands of evaluations of solutions, depending on the size of the design space and the complexity of the objective function. Direct optimisation of coronary stent design without surrogate model has been performed in literature [98, 116, 143], but these studies either use a small number of objectives [98], and/or perform low fidelity evaluations [116, 143]. Generally, the objective evaluations in optimisation often involve multiple computationally expensive CFD and/or FEM calculations. Surrogate modelling provides an effective alternative to performing these simulations. Once a surrogate model has been constructed using initial simulations, the optimisation algorithms can be coupled with surrogates for fast estimation of objective values across the design space and for all representation methodologies (size, shape, or topology). Surrogate assisted optimisation is the commonly used methodology for coronary stent designs (Table 4).

In terms of surrogate model types, various forms have been used in conjunction with optimisation algorithms. Polynomial regression and Gaussian process models (Kriging) [139] are common in stent optimisation literature (Table 4). Polynomial response surface models are the simplest surrogate models and usually produce satisfactory results [60, 83]. Gaussian process models have also been increasingly used [50, 66, 79, 123, 130], since these models can handle non-linear relationships. All these methods allow categorical variables such as material choices through one-hot encoding. In addition of creating surrogates for optimisation, it is vital to estimate the accuracy of these models. A common methodology involves the use of "leave-out" principle wherein either a single design point [53] or a randomised subset of multiple design points [50, 54] is left out during the creation of the surrogate to estimate the model's accuracy. This process is then repeated for all design points to estimate the parameters concerning the statistical accuracy of the surrogate model. While Standardised Cross Validation Residual (SCVR) [50, 53, 60] and R-squared ($R^2$) [60, 80, 127] are most common accuracy metrics, the use of other parameters such as mean squared error [54, 101] and t-values [60] is also reported. The surrogate-assisted optimisation process can be conducted directly on the surrogates created from initial sampling points [60, 80, 83, 101], or the surrogates can be updated periodically after true numerical evaluation of either a single [50, 54, 66, 86] or multiple [79] promising solutions, called infill point(s). The selection of these infill point(s) is based on an acquisition function which can be a solely based on predicted performance [66, 79] (believer model) or a composite function of predicted performance along with uncertainty such as Expected Improvement [50, 54, 66, 86] or Probability of Improvement [50]. A noteworthy application of Kriging believer model used NSGA-II algorithm to perform multi-objective multi-disciplinary optimisation while periodically updating surrogate model with five infill points in every iteration [79]. In another application of Kriging based optimisation[66], a single infill point with the maximum expected hypervolume improvement was evaluated in each iteration to highlight the benefits of uncertainty considerations in stent design optimisation.

Recent developments in ML have increased focus on Deep Neural Network (DNN) based surrogate models. These models have been successfully applied as surrogate to FEM simulation in biomedical applications such as the prediction of heart valve deformation [150]. However, the DNN surrogates are usually "black-box" models and do not yet have robust uncertainty metrics as compared to those provided by Kriging. This is a significant disadvantage for using these surrogate models in conjunction with optimisation algorithms that aim to maintain balance between the surrogate model uncertainty and objective minimisation while exploring the design space.

The current work focussed on coronary stent designs. However, the surrogate modelling and optimisation algorithms described above are equally applicable for different cardiovascular applications. These techniques were successfully used for single [151, 152], and multi objective [153, 154] optimisation of Nitinol based self-expandable stents used for femoral stenting applications. Additionally, these numerical techniques have been recently employed in design [155-157] and material property [158] optimisation of Nitinol based Transcatheter Aortic Valves.

### 4.4 Testing

Beyond the verification of the surrogate predictor as itself described above, once an optimal design has been found, it is recommended to assess the predicted design via testing to ensure superior performance relative to the baseline design. The testing of the derived optimal stent design can be undertaken using three different approaches: *in vivo, in vitro, and in silico. In vivo* testing in biological beings is the most comprehensive technique but is also time-consuming, costly, and presents ethical challenges. *In vitro* testing on the bench is a cost-effective technique and provides comprehensive analysis and visualisation opportunities, however, is limited in the ability to replicate all *in vivo* aspects including but not limited to deliverability, and deployment characteristics related to tissue behaviour etc. *In silico* methods relying on computational techniques are often comparatively more accessible, cost- and time effective. Yet, simplifying assumptions are required which demand careful considerations and results may only be interpreted within the realms of these assumptions of the mechanical behaviour of the tissue and stent involved. Thus it can be helpful to complement computational models with bench experiments to maximise the advantages of each approach [159, 160]. Different combinations of *in silico* [88, 89, 98, 100, 161], *in vitro* [88, 100, 143], and *in vivo* testing [88, 143] have been used to assess the performance of optimised designs. No optimisation study involves *in vitro* or *in vivo* assessment before final design testing because it would require creation of physical stents for initial sampling of the design space and for all iterations of optimisation algorithms. The most extensive final testing of optimised stent design involved *in silico* evaluation of dogboning, radial recoil, foreshortening, radial strength, and axial flexibility with additional *in vitro* evaluation of the two last metrics performed by Chen and co-authors [88]. This work used the micro-area x-ray diffraction for the first time to validate the residual stresses obtained after recoil in deployed stents with the virtual simulation. Final stage testing also included *in vivo* assessment rabbit models to measure BRS degradation. The extensive testing allows complete evaluation of optimised design to ensure no degradation is observed in other performance metrics that were not included in optimisation.

*In vitro* benchtop research also plays an important role in stent design development by comparing different designs for clinically relevant metrics such as longitudinal strength and stent fatigue. The designs with higher number of connectors (three versus two) showed significantly higher longitudinal resistance but compromised on bending flexibility [57, 91]. Additionally, the out-of-phase ring alignment improves longitudinal resistance due to ring-to-ring contact during compression [129]. These results forced changes in commercial stent design from two connector to three or four connector designs. For instance, the Promus Element stent was modified from purely two connector design to the Promus PREMIER (Boston Scientific Corp.) with four connectors at two proximal rings [57]. As for stent fatigue, designs with straight connectors and cobalt/platinum alloy material showed high fatigue resistance with no fracture reported at 10 million cycles in Platinum based Promus PREMIER and Integrity (Medtronic) verses fracture at less than hundred thousand cycles in curved connector of stainless-steel BioMatrix Flex BES (Biosensors International) design [102].

Optimisation process provides an algorithmic methodology to explore variable space for achieving improved designs, which is a scientific alternative to random search or human crafted searches. In addition to improving designs, surrogate assisted optimisation can immensely improve design cycle cost and speed. Our group performed stent design MOO across design space of seven geometry variables through surrogate models [66]. 30 numerical simulations (2 hours each on 48 core "Intel Cascade Lake Xeon" CPU cluster) were used to generate surrogate models for performing genetic optimisation. Similarly, neural networks based airfoil shape optimisation was recently performed within seconds compared to hours with CFD based traditional method [162]. This indicates that optimisation can have a large impact on costs and time associated with product design cycle and these efficiencies can be achieved in stent design cycles as well.

# 5 Key takeaway and future opportunities

The delineation of the impacts of geometry and drug-elution properties on stent performance was unlocked through population-based stochastic optimisation algorithms to date. However, some limitations persist as current optimisation methods are unable to search across different stent design families such as different helix-based and independent ring designs, optimise underrepresented clinically relevant and haemodynamic objectives such as overexpansion capacity, malapposition, and haemodynamic shear stress or secondary flow metrics, and evaluate the effect of non-geometric design features such as material properties. The means to accommodate these deficiencies is an area that needs further research attention.

First, the consideration of simultaneous search across multiple contemporary stent families by hybrid optimisation is an opportunity in the field. The design optimisation studies of commercial stents have mainly focussed on parametrising a baseline design and demonstrating the effects of changing design variables on performance metrics [54, 79, 80]. These perturbation-based shape optimisation studies parametrise single commercial independent ring design that results in high similarity between optimised and baseline designs. Optimisation studies are yet to tackle single (Resolute Onyx [21]) or a double helix design (Orsiro [20]). Moreover, it is critical to numerically evaluate and optimise across all possible design solutions including across different stent design families i.e., independent ring and helix-based designs, yet todays shape optimisation methods are not capable of this yet. Although the topology optimisation methods can search across stent design families, their computational demand and the current maximum number of three different simplified objectives, drastically limit their usefulness [132]. Machine learning tree based representation strategies are very promising due to their capacity of handling categorical variables while providing greater flexibility for representing multiple stent families. Specifically, Bayesian optimisation relying on Tree-Parzens Estimator [163] formulation can be exploited to iteratively identify a single infill solution across multiple commercial stent families.

Second, a comprehensive performance account of clinical relevance may drastically challenge our understanding of stent design success mechanisms. The stent design optimisation research has been overwhelmingly focussed on improving standard engineering metrics such as radial strength and axial flexibility. While a few clinically relevant metrics such as recoil and foreshortening have been optimised before, equally critical metrics have either been optimised only very recently such as longitudinal resistance [50, 93], or have not been used in design optimisation at all yet are daily clinical considerations including overexpansion capacity, and short as well as long-term malapposition. Instead, only *in vitro* studies of different stent designs were carried out to recommend design changes for higher longitudinal integrity [57, 91, 129] and provide overexpansion capacity of stents for clinical use [63, 64]. Malapposition has only been used to optimise dilation procedure [96] and not stent design till date. Moreover, haemodynamics play an important role in pathophysiological processes underlying vascular remodelling, disease processes and stenting success [49, 105, 164]. While shear stress related metrics such as WSS and TAWSS have been incorporated in optimisation studies [66, 86], the effect of related quantities such as Topological Shear Variation Index (TSVI) [165], RRT [108] and OSI [108] remain little defined together with recently emerging secondary flow consideration such as helicity on restenosis [75, 166] is yet to be fully studied . These can be a good candidate in optimisation studies for providing directions for future stent designs.

Third, stent design optimisation research has so far focussed on parametrising stent geometry. However, stent design performance is also dependent on other aspects such as strut- and connector base and coating materials. These materials and their properties affect different performance metrics such as mechanical, drug distribution, and biological response. In optimisation literature, materials have not been the focus. Only a few optimisation studies included factors that affect material properties, such as manufacturing parameters as a design variable [60, 81]. Additionally, material properties are even more critical for BRS, since dissolvable materials are inherently weaker. Most BRS-based optimisation studies tackle the issue of weaker BRS materials by focusing solely on structural objectives such as radial strength[83]. Including different material choices as a categorical design variable during optimisation is likely to lead to improved BRS designs.

Fourth, numerous performance metrics have been defined for optimisation of coronary stent design (Table 2). While a researcher may try to achieve a target design with zero recoil, the conflicting nature of performance metrics would lead to worse outcomes for other objectives such as arterial stresses [79]. Thus, an optimal strategy is to achieve Pareto optimal solutions, wherein the incremental design improvement is achieved for all objectives. Such an optimisation philosophy requires the researcher to have access to the expected value of all performance metrics of current generation of stent designs [50]. While these expected values are available in literature for some performance metrics (Table 2), efforts should be made to extract those for all metrices of contemporary stent designs.

Overall, stents have had a remarkable history and development to date, and it is difficult to predict were this field is destined to head overall. Persisting challenges remain including the 1) accurate description and variation of affected vascular and plaque tissue, and 2) consideration of a large number of at times competing performance objectives across the domains of structure mechanics, haemodynamic and practice application, and 3) the lack of ethical and realistic physical evaluation in the form of animal testing, one-off custom stent design manufacturing and realistic bench top replication, and 4) the availability of target values for all performance metrics of contemporary stent designs to be used in optimisation studies.

One hypothesis is that the large variation of patient lesion characteristics including arterial geometry, plaque geometry and stiffness, may benefit from an innovative custom approach beyond the currently available "off-the-shelf" stent designs from a small number of manufacturers - especially for highly complex and bifurcation stenting. Whilst dedicated bifurcation stents already exist- Tryton side branch stent (Tryton Medical, North Carolina, USA ) [167], continuous research and development in rapid image segmentation and reconstruction techniques [168], stent design optimisation, and additive manufacturing technology [169] may open up new pathways for custom designs. A recent work that developed personalised stent designs for various plaque configuration is a step in this direction [170]. Whilst the technology to support personalised stenting efforts may be well underway, we would like to highlight that this remains an unfeasible vision as of now because of significant regulatory and cost hurdles, together with the fact that a significant clinical benefit remains questionable on top of economical scale constrains since most current stent cases have remarkably good performance. One other noteworthy development presents PCI with drug coated balloons rather than stents, which recently shown promising results in specific patient cohorts with small vessel disease (vessel diameter < 3 mm) and high bleeding risk [171, 172]. These approaches and efforts may develop further to provide complementary or in some cases alternative solutions to stent implants in the future.

## 6 Conclusion

Coronary stents designs have undergone significant improvements in the past half-century in terms of both geometry and materials. The current industry consensus has evolved towards open cell independent ring or helix-based designs with metallic alloys such as cobalt chromium and platinum chromium for DES. Whilst these designs have successfully reduced restenosis and thrombosis events, further reduction will still significantly lower human hardship, considering thousands of coronary stent failures every year and promise to further translate into related aspects which are under-researched such as bifurcation or other complex lesion stenting. Especially multi objective optimisation accelerated stent design improvements to achieve an appropriate balance between the competing metrics across different domains including structural and haemodynamic, and considerations such as deliverability. Persisting challenges with scope for future improvement include the inability to search across different stent design families, underrepresentation of more clinically relevant and haemodynamic objectives, and the need to evaluate the impact of non-geometric design features such as material properties. These should be explored in parallel to alternative approaches, which as of today, either only have long-term or highly specialised applicability, with stents prevailing as preferred treatment of the largest social and economic health burden to date.

## 7 Acknowledgements

The authors acknowledge the support from competitive Australian National Heart Foundation Vanguard funding. A.K. further acknowledges the support from the Commonwealth Government through the Australian Government Research Training Program Scholarship and the computational cluster Katana supported by Research Technology Services at UNSW Sydney.

## 8 Contributions

A.K. performed the literature review, created figures, and wrote the manuscript. P.H.L., N.W.B, N.J., T.R., and S.B., contributed to the review and writing of the manuscript. T.R. and S.B. supervised the project.

## 9 Competing interests

The authors declare no competing interests.

# 11 Appendix

Table 5: Stent design optimisation studies categorisation based on performance metrics and geometric design variables.

| Metric type | Objective | Strut Width | Strut Length | Strut Thickness | Strut Cross Section | Strut Centreline shape | Strut Number per cell | Crown Radius | Connector Width | Connector Height | Connector Number per cell | Ring Number per stent | Ring Spacing |
|---|---|---|---|---|---|---|---|---|---|---|---|---|---|
| **Structural** — Intra-deployment | Recoil | Pant et al. [79], Amirjani et al.[55], Tammareddi et al. [80], Li et al. [81], Ribeiro et al. [50, 54], Li et al. [83] | Pant et al. [79], Amirjani et al.[55], Li et al. [81], Ribeiro et al. [50, 54], Li et al. [83] | Tammareddi et al. [80], Li et al. [81] | | | | Li et al. [83] | Amirjani et al.[55], Tammareddi et al. [80], Ribeiro et al. [50, 54] | Pant et al. [79], Ribeiro et al. [50, 54] | | | |
| | Arterial stresses (structural) | Pant et al. [53, 79], Amirjani et al.[55], Tammareddi et al. [80], Putra et al. [86], Ribeiro et al. [50, 54] | Pant et al. [53, 79], Amirjani et al.[55], Tammareddi et al. [80], Ribeiro et al. [50, 54] | Putra et al. [86] | | | | | Amirjani et al.[55], Tammareddi et al. [80], Ribeiro et al. [50, 54] | Pant et al. [53, 79], Ribeiro et al. [50, 54] | | | Timmins et al. [85], Putra et al. [86] |
| | Bending flexibility (crimped) | Pant et al. [53, 79], Ribeiro et al. [50, 54], Shen et al. [93] | Pant et al. [53, 79], Ribeiro et al. [50, 54], Shen et al. [93] | | | | | | Ribeiro et al. [54] [50], Shen et al. [93] | Pant et al. [53, 79], Ribeiro et al. [50, 54] | | | |
| | Stent stresses / strains (structural) | Wu et al. [87], Amirjani et al. [55], Blair et al. [60], Torki et al [89], Li et al. [83] | Amirjani et al.[55], Li et al. [83] | Torki et al [89] | | | Torki et al [89] | Torki et al [89], Li et al. [83] | Amirjani et al.[55] | | | Torki et al [89] | |
| | Longitudinal resistance | Ribeiro et al. [50, 54], Shen et al. [93] | Ribeiro et al. [50, 54], Shen et al. [93] | | | | | | Ribeiro et al. [50, 54], Shen et al. [93] | Ribeiro et al. [50, 54] | | | |

| Category | | Metric | | | | | | | | | | | |
|---|---|---|---|---|---|---|---|---|---|---|---|---|---|
| | | Dogboning | Li et al. [125], Ribeiro et al. [54] | Ribeiro et al. [54] | Li et al. [125] | | | | | | Ribeiro et al. [54] | Ribeiro et al. [54] | |
| | | Foreshortening (%) | Li et al. [81], Blair et al. [60], Torki et al [89] | Li et al. [81], Blair et al. [60], | Li et al. [81], Blair et al. [60], Torki et al [89] | | Torki et al [89] | Torki et al [89] | | | | | Torki et al [89] |
| structural | Post-deployment | Radial strength | Ribeiro et al. [50, 54], Li et al. [83] | Ribeiro et al. [50, 54], Li et al. [83] | Baradaran et al. [98] | | Baradaran et al. [98] | | Li et al. [83] | Ribeiro et al. [50, 54] | Ribeiro et al. [50, 54] | | |
| | | Collapse pressure | Blair et al. [60] | Blair et al. [60] | Blair et al. [60] | | | | | | | | |
| | | Radial stiffness | Grogan et al. [97], Clune et al. [101], Gharleghi et al. [66] | Grogan et al. [97] | Grogan et al. [97], Gharleghi et al. [66] | Gharleghi et al. [66] | Clune et al. [101] | | | | | Gharleghi et al. [66] | Gharleghi et al. [66] |
| | | Bending flexibility (deployed) | | | Baradaran et al. [98] | | Baradaran et al. [98] | | | | | | |
| | | Fatigue resistance | Clune et al. [101], Ribeiro et al. [54] | Ribeiro et al. [54] | | | Clune et al. [101] | | | Ribeiro et al. [54] | Ribeiro et al. [54] | | |
| | | Stent artery coverage | Blair et al. [60], Li et al. [83] | Blair et al. [60], Li et al. [83] | Blair et al. [60], | | | | Li et al. [83] | | | | |
| Haemodynamic | Shear Stress | Wall Shear Stress (WSS) | Amirjani et al.[55], Putra et al. [86] | Amirjani et al.[55] | Putra et al. [86] | Gharleghi et al. [66] | | | | Amirjani et al.[55] | | Gharleghi et al. [66] | Putra et al. [86], Gharleghi et al. [66] |
| | | Time Average WSS (TAWSS) | Pant et al. [79] | Pant et al. [79] | | Gharleghi et al. [66] | | Gundert et al. [114] | | | Pant et al. [79] | Gharleghi et al. [66] | Gharleghi et al. [66] |

| | | | | | | |
|---|---|---|---|---|---|---|
| **Haemodynamic** Secondary flow | Secondary flow metrics – Helicity/ 2D vorticity / Recirculation zone length | Srinivas et al. [113] | Srinivas et al. [113] | | | Srinivas et al. [113] |
| **Combined** | Tissue prolapse | Ribeiro et al. [50, 54] | Ribeiro et al. [50, 54] | | Ribeiro et al. [50, 54] | Ribeiro et al. [50, 54] |
| | Stent drug elution | Pant et al. [53, 79] | Pant et al. [53, 79] | | | Pant et al. [53, 79] |